\documentclass[%
reprint,
 amsmath,amssymb,
 aps,showkeys,
pre
]{revtex4-1}

\usepackage{graphics}
\usepackage{graphicx}
\usepackage{dcolumn}
\usepackage{bm}
\usepackage{hyperref}
\usepackage{placeins}

\newcommand\erfc{\mbox{erfc}}

\begin{document}

\title{Synchronization transition in ensemble of coupled phase oscillators with coherence-induced phase correction }
\date\today

\author{S. Belan}
\affiliation{Moscow Institute of Physics and Technology, 141700 Dolgoprudny, Russia}
\affiliation{Landau Institute for Theoretical Physics,
 142432 Chernogolovka, Russia}

 \begin{abstract}

We study synchronization phenomenon in a self-correcting population of noisy  phase oscillators with  randomly
distributed natural frequencies.
In our model  each  oscillator stochastically switches its phase to the ensemble-averaged value $\psi$ at a typical rate which is proportional to the degree of phase coherence $r$.
The system exhibits a continuous phase transition to collective synchronization similar to classical Kuramoto model. 
Based on the self-consistent arguments and on the linear stability analysis of an incoherent state we derive analytically the threshold value $k_c$ of coupling constant corresponding to the onset of a partially synchronized state.
Just above the transition point the linear scaling law $r\propto k-k_c$ is found.
We also show that nonlinear relation between  rate of phase correction and order parameter leads to non-trivial transition between incoherence and synchrony. 
To illustrate our analytical results, numerical simulations have been performed for a large population of phase oscillators with proposed type of coupling.
The results of this work could become useful in designing distributed networked systems capable of self-synchronization.

 \end{abstract}


\maketitle

\section{Introduction}

The  spontaneous synchronization of mutually coupled oscillators is observed in many complex biological, chemical, physical and sociological systems with different origins of rhythmical activity and different mechanisms of coupling.
Depending on the context the term "oscillator" may refer to neurons \cite{Hopfield_1995}, cardiac pacemaker cells \cite{Peskin_1975},  yeast cells \cite{Ghosh_1971}, flashing fireflies \cite{Buck_1976},  chirping crickets \cite{Sismondo_1990}, applauding spectators \cite{Nikitin2001}, nano-mechanical resonators \cite{Bagheri_2003}, lasers \cite{Jiang_1993}, superconducting Josephson junctions \cite{Wiesenfeld_1996},
 etc.
The concept of spontaneous synchronization has proved to be useful in designing of artificial networked systems capable of self-organization in the absence of any centralized control mechanism \cite{Arenas_2008}.
A common example of these systems is the so-called wireless sensor networks that typically consist of a number of spatially distributed devices with computational, sensing, memory and wireless communication capabilities \cite{Hekmat_2006}.
Many industrial and civil applications of sensor networks require global clock synchronization of sensors \cite{Wu_2011} or their distributed consensus on certain quantities of interest \cite{Olfati-Saber_2004}. 
A good synchronization scheme should provide the robustness with respect to
failure of an individual sensors and take into account the diversity of sensor's properties and the presence of noise. 
The nature-inspired approach to this problem is completely distributed synchronization strategies based on the mutual coupling between sensors \cite{Simeone_2007,Carli_2008,Scutari_2008,Schenato_2011,Monti_2014}.   
In self-organizing sensor network the nodes are coupled through the exchange of information:  each node transmit its own state and   receive the signals from the neighbors.
The knowledge of this information 
allows to single node to update itself occasionally in accordance with some correction rule so as to facilitate the global synchrony.
Another important technological application of synchronization phenomenon lies in the field of smart power grids \cite{Rohden_2012,Motter_2013}.
The modern electrical grids consist of thousands of generators and substations linked across large distances.
To ensure stable operation and efficiency of the entire system, all the components that generate electricity must operate at the same frequency.
In a decentralized smart power grid, the nodes could communicate among each other and correct their parameters in order to achieve and maintain the synchronized state.

Motivated by the discussed applications, in this work, we investigate an ensemble of phase oscillators for which the mutual coupling has a form of phase correction.
Our model implies that each oscillator evolves autonomously except for discrete times when the instantaneous correction events occur.
We focus our attention on the simple correction rule which prescribe to oscillator to reset its phase to the ensemble-averaged value at a typical rate which is  determined by the product of the coupling constant
and the degree of global phase coherence. 
This basic model is inspired by the famous Kuramoto model in which the effective coupling force is proportional to the amplitude of the mean field \cite{Kuramoto_1975}.
The central result of the work is a synchronization transition upon the change of coupling constant.
We demonstrate that for sufficiently strong coupling a partially synchronized state continuously bifurcates from incoherence.
It is also shown how the synchronization properties of oscillators change in the case of non-linear relation between the correction rate and the degree of synchrony.

\section{Model formulation}

Consider a diverse ensemble of $N$ noisy oscillators which are solely characterized by the phase variables $\varphi_i\in [-\pi,\pi]$, where $1\le i\le N$.
We will use the  Kuramoto's order parameter  \cite{Kuramoto_book,Kuramoto_1975} 
\begin{equation}
\label{order parameter0}
re^{i\psi}=\frac{1}{N}\sum\limits_{j=1}^{N} e^{i\varphi_j},
\end{equation}
 to describe the collective rhythm produced by the whole system.
  The magnitude $0\le r\le 1$ of this complex parameter measures the degree of macroscopic coherence, while $\psi$  is the average phase.
In the absence of coupling the dynamics of the $i$th oscillator is given by  
\begin{equation}
\label{phase dynamics}
\partial_t\varphi_i=\Omega_i+\xi_i,
\end{equation}
where  the intrinsic frequency $\Omega_i$ is randomly chosen from some probability density $g(\Omega)$ 
and $\xi_i(t)$ is  the Gaussian white noise,
\begin{equation}
\langle\xi_i(t)\rangle=0,\ \ \ \langle\xi_i(t_1)\xi_j(t_2)\rangle=2D\delta_{ij}\delta(t_1-t_2).
\end{equation}
The the angular brackets denote averages over statistics of the noise and the non-negative parameter $D$ measures its intensity. 
Dropping the index, we introduce the one-oscillator probability density of a phase distribution as $n(\varphi,\Omega,t)=\langle\delta(\varphi-\varphi(t)) \rangle$,
 where $\varphi(t)$ is a solution of the stochastic equation (\ref{phase dynamics}) for a fixed realization of noise. 
The time evolution of this function obeys the 
following Fokker-Planck equation \cite{Risken} 
\begin{equation}
\label{FPE0}
\partial_t n=D\partial_\varphi^2 n-\Omega\partial_\varphi n.
\end{equation}
Note that $n$ is nonnegative, $2\pi$-periodic in $\varphi$, and satisfies
the normalization condition 
\begin{equation}
\label{normalization condition}
\int\limits_{-\pi}^{+\pi} n(\varphi,\Omega,t)d\varphi=1.
\end{equation}
In the thermodynamic limit $N\to\infty$, the order  parameter (\ref{order parameter0})  can be expressed as
\begin{equation}
\label{order parameter1}
re^{i\psi}=\int\limits_{-\infty}^{+\infty}d\Omega g(\Omega)\int\limits_{-\pi}^{\pi} e^{i\varphi}n(\varphi,\Omega,t)d\varphi.
\end{equation}

For an arbitrary initial distribution $n(\varphi,\Omega,0)$ the solution of Eq. (\ref{FPE0}) is given by the convolution  $n(\varphi,\Omega,t)=\int_{-\pi}^{\pi}G(\varphi-\varphi',t)n(\varphi',\Omega,0)d\varphi'
$, 
where the Green function
\begin{equation}
\label{green function}
G(\varphi,\Omega,t)=\frac{1}{2\pi}+\frac{1}{\pi}\sum\limits_{m=1}^{\infty}e^{-m^2Dt} \cos m(\varphi-{\Omega t}),
\end{equation}
obeys the Fokker-Plank equation (\ref{FPE0}) together with the initial condition $G(\varphi,\Omega,0)=\delta(\varphi)$. 
It follows from (\ref{green function}) that any initial state  relaxes exponentially fast to the uniform distribution  $n(\varphi,\Omega)=1/2\pi$, which corresponds to zero order parameter (\ref{order parameter1}).
Predictably, the diverse population of non-interacting noisy units behaves incoherently.

Next, let us assume that oscillators are coupled and update itself in order to improve the degree of global synchrony. 
Namely, the $i$th oscillator evolves accordingly to the equation (\ref{phase dynamics}),
but from time to time it instantaneously shifts the phase to the current ensemble-averaged value $\psi$. 
We treat the updating time instants as completely random and
statistically independent for different oscillators.
The intensity of phase correction is characterized by a typical rate of updates $\alpha$ per oscillator which, in general, may change over time. 
Under these assumptions, the phase dynamics can be viewed as a drifting diffusion process on a circle in the presence of stochastic resetting \cite{Evans_2011}  with $\psi$ and $\alpha$ playing a role of resetting position and resetting rate, respectively.
We, thus, write the following master equation in the limit $N\to\infty$
\begin{equation}
\label{FPE2}
\partial_t n=D\partial_\varphi^2 n-\Omega\partial_\varphi n-\alpha n+\alpha\delta(\varphi-\psi),
\end{equation}
where $\psi$ is determined by (\ref{order parameter1}) and the one-oscillator probability density $n(\varphi, \Omega,t)$ implies the averaging of phase dynamics over both the noise and stochastic phase correction. 
The structure of the equation (\ref{FPE2}) is quite transparent: the first and second terms in the right hand side correspond to a diffusion and uniform rotation at natural frequency (cf. with Eq. (\ref{FPE0})), while the third and fourth terms are related to updates and represent a negative
probability flux  out of each point $\varphi$ and a
corresponding positive probability flux into $\varphi=\psi(t)$.
Taking into account the $2\pi$-periodicity of function $n$ in $\varphi$ it is easy to show the conservation of total probability (\ref{normalization condition}).

The equation (\ref{FPE2}) constitutes the basis of our statistical model for coupled oscillators.
The focus of this study is the synchronization properties of the system in the case when the correction rate $\alpha$ is a function of the mean-field amplitude $r$. 
It is useful to start with the description of the steady-state dynamics arising from the Eq. (\ref{FPE2})  with constant parameter $\alpha$.




\section{Steady-state analysis}

Suppose that the rate of updates $\alpha$ is time-independent.
Since the phase correction promotes the consistent behaviour of oscillators, the incoherent state $n(\varphi,\Omega)=1/(2\pi)$ does not solve the Eq. (\ref{FPE2}) at $\alpha\ne 0$.
If frequency distribution $g(\Omega)$   has a single maximum at frequency $\Omega_0$ and is symmetric, it is reasonable to seek  $n(\varphi,\Omega,t)$ in the form of some stationary profile uniformly rotating at frequency $\Omega_0$. 
Then passing into the rotating frame $\varphi\to \varphi+\Omega_0 t$ and neglecting the time fluctuations of the order parameter at $N\to\infty$ one can set $\psi=0$ without loss of generality. 
In result, the problem reduces to the stationary equation
\begin{equation}
\nonumber
D\partial_\varphi^2 n-\omega\partial_\varphi n-\alpha n+\alpha\delta(\varphi)=0,
\end{equation}
 supplemented by the periodic condition $n(-\pi,\omega)=n(\pi,\omega)$, where $\omega=\Omega-\Omega_0$. 
It is straightforward to find the following solution
\begin{equation}
\label{solution}
n(\varphi,\omega)=\left\{\begin{array}{ll}
A_{1} \exp[\gamma_1\varphi]+A_{2} \exp[\gamma_2\varphi] ,\ \ 0\le \varphi\le \pi,\\
\\
B_{1} \exp[\gamma_1\varphi]+B_{2}\exp[\gamma_2\varphi],\ \ -\pi<\varphi\le 0,
\end{array} \right.
\end{equation}
in which $\gamma_{1,2}=(\omega\pm\sqrt{\omega^2+4\alpha D})/2D$ 
 and 
 \begin{eqnarray}
A_{1,2}=\frac{\alpha}{(e^{2\pi\gamma_{1,2}}-1)\sqrt{\omega^2+4\alpha D}},\\
\label{coefficients}
B_{1,2}=\frac{\alpha e^{2\pi\gamma_{1,2}}}{(e^{2\pi\gamma_{1,2}}-1)\sqrt{\omega^2+4\alpha D}}.
\end{eqnarray}
It is noteworthy, that for $\alpha\ne 0$ this stationary distribution is nonequilibrium since phase correction produces ongoing circulation of probability.  

We can now calculate the order parameter (\ref{order parameter1}) by using (\ref{solution}-\ref{coefficients}).
Since $\psi=0$  by assumption, one readily obtains
\begin{equation}
\label{order_parameter5}
r=\alpha(\alpha+D)\int\limits_{-\infty}^{+\infty}d\omega\frac{g(\Omega_0+\omega)}{\omega^2+(D+\alpha)^2}.
\end{equation}
In a particular case of a Lorentzian distribution $g(\Omega)=(\gamma/\pi)(\gamma^2+(\Omega-\Omega_0)^2)$ we find exactly 
\begin{equation}
\label{Lorentzian}
r=\frac{\alpha}{\alpha+D+\gamma},
\end{equation}
whereas for a normal distribution with standard deviation $\sigma$ one obtains
\begin{equation}
\label{normal}
r=\frac{\alpha}{\sqrt{2\pi}\sigma}\exp\left(\frac{(D+\alpha)^2}{2\sigma^2}\right)\erfc\left(\frac{D+\alpha}{\sqrt2\sigma}\right),
\end{equation}
where $\erfc$ denotes the complementary error function.


Importantly, the result (\ref{order_parameter5}) is extracted from the equation (\ref{FPE2}) which is justified by the assumption that updates are the random Poisson events.
There is a more general way to derive the steady-state coherence $r$ which is applicable to the case of arbitrary stationary statistics of updating time instants as long as these instants are independent for different oscillators.
As previously, we suppose that density profile of oscillators in the phase circle is stationary in the the rotating frame, $\varphi\to \varphi+\Omega_0 t$, and set $\psi=0$.  
Then the oscillator dynamics may be thought of as a renewal process: each update resets the phase of oscillators to $\varphi=0$, whereas in the inter-update time intervals the evolution of phase is given by superposition of diffusion and detuning drift. 
The Green function (\ref{green function}) rewriting in the rotating frame yields the probability density of phase distribution for a randomly chosen oscillator 
\begin{equation}
\label{green function rotating}
G(\varphi,\omega,\tau)=\frac{1}{2\pi}+\frac{1}{\pi}\sum\limits_{m=1}^{\infty}e^{-m^2D\tau}\cos m(\varphi-\omega \tau),
\end{equation}
where $\tau$ denotes the time elapsed since the last update of this oscillator.
To obtain the steady-state distribution $n(\varphi,\omega)$ one should average the Eq. (\ref{green function rotating})  over different realization of the phase correction process, i.e. 
\begin{equation}
\label{stationary}
n(\varphi,\omega)=\int\limits_{0}^{+\infty}P(\tau)G(\varphi,\omega,\tau)d\tau,
\end{equation}
where $P(\tau)$ is the probability density of random variable $\tau$. 
After the substitution the expressions (\ref{green function rotating}), (\ref{stationary}) and $\psi=0$ into (\ref{order parameter1}) we derive the following simple formula
\begin{equation}
\label{order parameter4}
r=\int\limits_{-\infty}^{+\infty}\int\limits_{0}^{+\infty}d\omega d\tau g(\omega+\Omega_0)  P(\tau)e^{-D\tau}\cos\omega \tau,
\end{equation}
which allows us to readily determine the stationary coherence $r$ once $P(\tau)$ is  known. 
For  stochastic phase correction with Poisson rate of updates $\alpha$ we have $P(\tau)=\alpha e^{-\alpha \tau}$.
Then integration of (\ref{order parameter4}) over $\tau$ yields exactly (\ref{order_parameter5}).

As an example, let us apply the Eq. (\ref{order parameter4}) to the case of independent periodical correction processes.
We assume that the sequence of updating events $\{t_i(k)\}_{k=0}^\infty$ of $i$th oscillator is given by $t_i(k)=\tau_i+kT$, where $T$ is updating period and $\tau_i$ is a random variable uniformly distributed over the interval $[0,T]$.
Then  $P(\tau)=T^{-1}$ for $0\le\tau\le T$ and $P(\tau)=0$ otherwise.
Performing integration in (\ref{order parameter4}) with $g(\Omega)=(\gamma/\pi)(\gamma^2+(\Omega-\Omega_0)^2)$ one obtains
\begin{equation}
\label{Lorentzian+periodic}
r=\frac{1-e^{-(D+\gamma)T}}{(D+\gamma)T}.
\end{equation}
In agreement with intuition, Eqs. (\ref{Lorentzian}), (\ref{normal}) and (\ref{Lorentzian+periodic}) give $r\to 0$ (incoherence) and  $r\to 1$ (perfect synchrony) at $\alpha,T^{-1}\to 0$ and $\alpha,T^{-1}\to \infty$ respectively.


\section{State-dependent correction rate}

We now turn to the more non-trivial case when the correction rate $\alpha$ is not fixed but depends itself on the state of population.
Let us assume that updates occur the more frequently, the more higher the current degree of phase synchrony $r$.
Apparently, this coupling scenario
sets up a positive feedback so that as the population becomes more coherent, $r$ grows and so the correction rate $\alpha$ increases, which tends to synchronize the oscillators even stronger.
The situation is quite similar to classical Kuramoto model where the mean-field coupling becomes stronger with the growth of global synchrony.
It is well-known that Kuramoto-like coupled oscillators  exhibit spontaneous synchronization provided the interaction constant is large enough. 
The major question we wish to address here is whether there is a similar phenomenon in the discussed case of coupling through the coherence-induced phase correction.
 The answer is positive, below we demonstrate that for sufficiently strong coupling the phase correction compensates the destructive effects of noises and frequency’s diversity and a partially
synchronized state emerges.
We focus our attention on the linear relation between coherence and correction rate, i.e.  $\alpha=kr$, where $k\ge 0$ is the interaction constant. 
The nonlinear generalizations of the model will be discussed briefly in Section \ref{sec: nonlinear}. 



\subsection{Self-consistency analysis and numerics}

One may expect that starting from an arbitrary initial condition the population of oscillators will settle into a steady state with constant coherence $r$ and average phase $\psi$ running at frequency $\Omega_0$.
Then the correction rate $\alpha$ becomes time-independent and the results of the previous section can be straightforwardly applied.
Replacing $\alpha$ by $kr$ in Eq. (\ref{order parameter4}), we find the following self-consistency condition  for stationary coherence $r$
\begin{equation}
\label{self-consistent}
r=kr(kr+D)\int\limits_{-\infty}^{+\infty}d\omega\frac{g(\Omega_0+\omega)}{\omega^2+(D+kr)^2}.
\end{equation}
A trivial zero solution $r=0$, corresponding to a incoherent state, holds for any value of $k$.
A second brunch of solutions  bifurcates continuously at critical coupling 
\begin{equation}
\label{critical coupling}
k_c=\frac{1}{D}\left[\int\limits_{-\infty}^{+\infty}d\omega\frac{g(\Omega_0+\omega)}{\omega^2+D^2}\right]^{-1},
\end{equation}
obtained by letting $r\to+0$, and corresponds to a partially synchronized state with non-zero $r$.

Remarkably, formula (\ref{critical coupling}) has the same structure as that for the critical coupling in the noisy Kuramoto model \cite{Sakaguchi_1988}. 
At the same time, the models differ by the critical behaviour of the order parameter. 
An expansion of the right-hand side of Eq. (\ref{self-consistent}) in powers of $r$ yields the linear scaling law  $r\propto k-k_c$ in the vicinity of transition.
In contrast, in  Kuramoto model the order parameter of the bifurcating branch  obeys the square-root law $r\propto\sqrt{ k-k_c}$ near  threshold.
Note also that how $r$ grows with $k$ strictly depends on the statistics of updates.
For the sake of illustration, suppose that $g(\Omega)$ is a Lorentzian distribution.
Then, for stochastic phase correction we find from (\ref{Lorentzian}) exactly $r=(k-k_c)/k$ at $k\ge k_c=D+\gamma$.
However, for periodic phase correction with coherence-dependent frequency of updates $1/T(r)=kr$, one obtains from (\ref{Lorentzian+periodic}) the non-trivial solution $r= -(k/k_c)\ln^{-1}(k/k_c-1)$ at $k\ge k_c=D+\gamma$.

The emergence of spontaneous synchronization is confirmed by simulations of large but finite population of oscillators.
Our numerical scheme use temporal discretization of Eq. (\ref{phase dynamics})  with time step $\Delta t$ so that $\varphi_i^n$ is the approximation of $\varphi_i(n\Delta t)$, where $n=0,1,2,\dots$.
Then, the complex order parameter is $r_ne^{i\psi_n}=(1/N)\sum_{j=1}^{N}e^{i\varphi_j^n}$.
The random forces $\xi_i$ are modeled by the telegraph processes whose correlation times are equal to time step duration $\Delta t$.
Specifically, the values of $\xi_i$ inside the $n$th time step are chosen to be independent random constants $\xi_i^n$ with identical normal distributions.
To ensure a given value of the diffusion coefficient $D$, one
should choose $\langle \xi_i^{n2}\rangle=2D/\Delta t$.
At each time step we generate the set of $N$ random numbers $\tau_1^n,\tau_2^n,\dots,\tau_N^n$ having distribution $\alpha_n e^{-\alpha_n\tau}$ with $\alpha_n=kr_n$.
If $\tau_i^n\le\Delta t$, we switch the phase of $i$th oscillator to the current ensemble-averaged value $\psi_n$. 
Thus, the rules for the evolution of the system are as follows
\begin{equation}
\label{numerics}
\nonumber
\varphi_i^{n+1}=\left\{\begin{array}{ll}
\varphi_i^n + (\Omega_i + \xi_i^n)\Delta t,\ \  \text{if}\ \  \tau_i^n>\Delta t,\\
\\
\psi_n + (\Omega_i + \xi_i^n)(\Delta t-\tau_i^n), \ \  \text{if}\ \  \tau_i^n\le\Delta t.
\end{array} \right.
\end{equation}

\begin{figure}[t]
  \center{\includegraphics[scale=0.39]{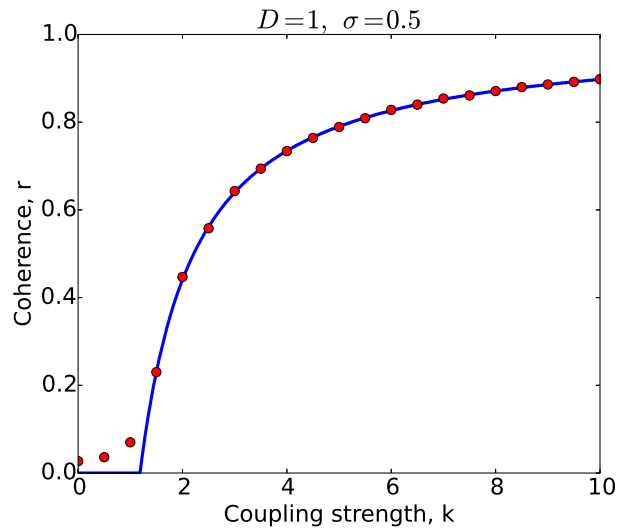}}
  \caption{The degree of phase coherence $r$ in dependence on coupling strength $k$ for population of $N$ noisy oscillators with  normal distribution $g(\Omega)$ of natural frequencies. The solid line shows the solution of the self-consistency equation (\ref{self-consistent}) corresponding to thermodynamic limit $N\to \infty$.  Numerical results are shown as points, and are taken from the simulations of $N=1000$ oscillators.  The simulations
were performed for $2 \times 10^5$ time steps, starting from uniform initial distribution of phases over the interval $[-\pi,\pi]$. The values presented in the figure are averages over the
last $1.8 \times 10^5$ steps.}
  \label{pic:coherence}  
  \end{figure}

In numerics the initial phases of $N=1000$ oscillators  were uniformly distributed over the interval $[-\pi,\pi]$
and the discrete time step $\Delta t=0.001$ was used.
The diffusivity was $D=1$ and the natural frequencies $\Omega_i$ were chosen from normal distribution $g(\Omega)$ 
with  expected value $\Omega_0=0$ and standard deviation $\sigma=0.5$.
Figure~\ref{pic:coherence} represents numerical results for the time-averaged coherence $r$ as a function of coupling constant $k$ in comparison with theoretical curve for infinite-$N$ system.
 It can be seen that level of synchrony in population remains low 
 until $k$ reaches a critical value $k_c\approx 1.2$, above which $r$ monotonically increases towards its asymptotic value of unity. 
The non-zero values of $r$ below $k_c$ in the simulation  
reflect fluctuations due to finite $N$.

\subsection{Global dynamics in homogeneous system}

The self-consistent arguments presented above do not provide any information about the stability of incoherent and partially synchronized states.
The direct analysis of the stability properties is possible when the  diversity of natural frequencies is negligible, i.e. $g(\Omega)=\delta(\Omega-\Omega_0)$.  
Then, using Eq. (\ref{FPE2}) one can derive the following closed-form equations associated with angular and radial motions of the
order parameter $re^{i\psi}=\int_{-\pi}^{+\pi}  e^{i\varphi}n(\varphi,t)d\varphi$
\begin{eqnarray}
\label{phase equation}
\dot{\psi}=\Omega_0,\\
\label{amplitude equation}
\dot{r}=-Dr-\alpha r+\alpha.
\end{eqnarray}
The equation (\ref{phase equation}) suggests that the average phase just uniformly rotates at frequency $\Omega_0$. 
Substituting $\alpha=kr$ we find easily the solution  of amplitude equation (\ref{amplitude equation}) 
\begin{eqnarray}
r(t)=\frac{(k-D)e^{(k-D)t}r_0}{k-D-k(1-e^{(k-D)t})r_0},
\end{eqnarray}
where $r_0=r(0)$ is an initial condition.
For $k<D$ the coherence $r$ always decays to zero as $t\to\infty$, but for $k>D$ incoherent state loses stability and one obtains another attracting point $r=(k-D)/k$ corresponding to partially synchronized state.
Since $g(\Omega)=\delta(\Omega-\Omega_0)$ by assumption, the critical value $k_c=D$ is consistent with (\ref{critical coupling}).

If the natural frequencies of oscillators are non-identical, there is no closed description of global dynamics in terms of the order parameter. 
In the next subsection we propose perturbative approach which allows us to demonstrate that incoherent solution  becomes linearly unstable as the strength of the coupling goes beyond the threshold (\ref{critical coupling}).

\subsection{Linear stability analysis of incoherence} 
  
In what follows we look at the oscillator dynamics in the
frame moving with frequency $\Omega_0$. 
Let us consider a small perturbation of the uniform incoherent state 
\begin{equation}
\label{perturbation}
n(\varphi,\omega,t)=\frac{1}{2\pi}+\varepsilon \rho(\varphi,\omega,t),
\end{equation}
where $\varepsilon\ll 1$.
At lowest (linear) order in $\varepsilon$ the evolution of perturbation is governed by equation
\begin{equation}
\label{FPE3}
\partial_t \rho=D\partial_\varphi^2 \rho-\omega\partial_\varphi \rho-\frac{kr}{2\pi}+kr\delta(\varphi-\psi).
\end{equation}
To proceed  we write a $2\pi$-periodic function $\rho$ as a superposition of Fourier harmonics 
\begin{equation}
\label{Fourier harmonics}
\rho(\varphi,\omega,t)=\sum\limits_{m=1}^{\infty}c_m(\omega,t)e^{im\varphi}+\sum\limits_{m=1}^{\infty}c_m^\ast(\omega,t)e^{-im\varphi}.
\end{equation}
Note that normalization condition (\ref{normalization condition}) automatically provides $c_0=c_0^\ast=0$.
When (\ref{Fourier harmonics}) is substituted into (\ref{FPE3}), we obtain the following 
equation for the amplitude $c_m$
\begin{equation}
 \label{high harmonics}
\partial_tc_m=-(m^2D+im\omega)c_n+kr(t)e^{-im\psi(t)}.
\end{equation}
 The evolution equation for $c_m^\ast$ is just the conjugate of (\ref{high harmonics}).
 
 Next, substituting (\ref{Fourier harmonics}) into (\ref{order parameter0}) yields
\begin{equation}
\label{order parameter}
re^{i\psi}=2\pi \int\limits_{-\infty}^{+\infty}d\omega g(\omega+\Omega_0) c_1^\ast(\omega,t).
\end{equation}
 Thus, only the first harmonic of the Fourier decomposition,  which is the so-called fundamental mode, contributes to the order parameter.

From (\ref{high harmonics}) and (\ref{order parameter}) one easily sees, that equation for the amplitude $c_1$ (and $c_1^\ast$) is uncoupled 
 \begin{equation}
 \label{fundamental mode}
 \partial_t c_1=-(D+i\omega)c_1+k  \int\limits_{-\infty}^{+\infty}d\nu g(\nu+\Omega_0) c_1(\nu,t).
 \end{equation}
 Let $c_1(\omega,t)=a(\omega)e^{\lambda t}$, then we obtain $\hat L a=\lambda a$, where $\hat L=-(D+i\omega)+k  \int_{-\infty}^{+\infty}d\nu g(\nu+\Omega_0) $.
  The spectrum of the operator $\hat L$ was constructed in \cite{Strogatz_1991} in the context of noisy  Kuramoto model \cite{Sakaguchi_1988}.
  They show that the  continuous part of spectrum lies in left half-plane for any $k$, thus corresponding modes never cause instability. 
  In contrast, the location of discrete spectrum depends strongly on interaction constant $k$. 
When $k$ exceeds a threshold (\ref{critical coupling}), the discrete real eigenvalue $\lambda>0$ appears so that incoherent state becomes unstable. 

Now we turn to the analysis of high order amplitudes.
The equation (\ref{high harmonics}) may be solved easily in
terms of $r(t)$, $\psi(t)$ and the initial condition $c_m(\omega,0)$.
The result is
\begin{eqnarray}
\nonumber
c_m(\omega, t)=c_m(\omega,0)e^{-(m^2D+im\omega)t}+\\
\nonumber
+ke^{-(m^2D+im\omega)t}\int\limits_0^tdt'e^{(m^2D+im\omega)t'}r(t')e^{-im\psi(t')}.
\end{eqnarray}
The order parameter is completely determined by the fundamental mode through Eq. (\ref{order parameter}), thus we put $r(t)e^{i\psi(t)}=r(0)e^{i\psi(0)+\lambda t}$, where the eigenvalue $\lambda$ belongs to the spectrum described above.
After integration one obtains
\begin{eqnarray}
\nonumber
c_m(\omega, t)=\frac{kr(0)e^{-im\psi(0)}}{m^2D+\lambda'+im(\omega-\lambda'')}e^{\lambda' t-im\lambda''t}+\\
\label{high amplitude}
+\left(c_m(\omega,0)-\frac{kr(0)e^{-im\psi(0)}}{m^2D+\lambda'+im(\omega-\lambda'')}\right)e^{-(m^2D+im\omega)t},
\end{eqnarray}
where $\lambda'$ and $\lambda''$ are the real and imaginary parts of $\lambda$, respectively.
Notice that the denominator in (\ref{high amplitude}) never vanishes since $m> 1$ and $\lambda'\ge-D$ for any $k$ \cite{Strogatz_1991}.
The second term in the rhs of (\ref{high amplitude}) rapidly dies out, while the behaviour of the first one depends on the coupling strength $k$.
At $k<k_c$ the spectrum $\lambda$ lies in the left half-plane so that $c_m(t)$ always decays to zero. 
For $k>k_k$ the aforementioned positive discrete eigenvalue provides the exponential growth of  $c_m(t)$.

Let us summarize the results of this subsection. 
The incoherent state  goes linearly unstable to perturbation involving first harmonic for $k>k_c$, where the critical coupling $k_c$ coincides with the prediction (\ref{critical coupling}) of self-consistency analysis.
Above the threshold the instability of the first harmonic induces also the growth of the high order amplitudes.

%
%

\section{ Nonlinear generalization}
\label{sec: nonlinear}
  
  \begin{figure}[t]
  \center{\includegraphics[scale=0.36]{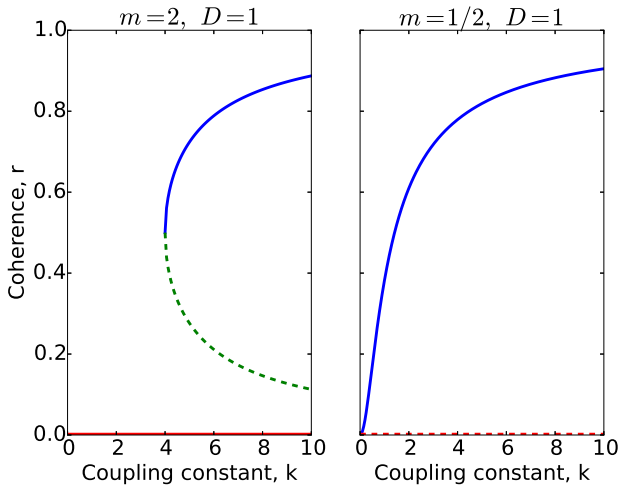}}
  \caption{The phase coherence $r$ in dependence on the coupling strength $k$ for infinite ensemble of identical noisy oscillators with nonlinear phase correction rate $\alpha=k r^m$, where $m=2$ (left subplot) and $1/2$ (right subplot). The diffusion coefficient is $D=1$. The solid and the dashed lines correspond to the stable and unstable branches respectively.
     }
  \label{pic:nonlinear}  
  \end{figure}
  
\begin{figure}[t]
  \center{\includegraphics[scale=0.37]{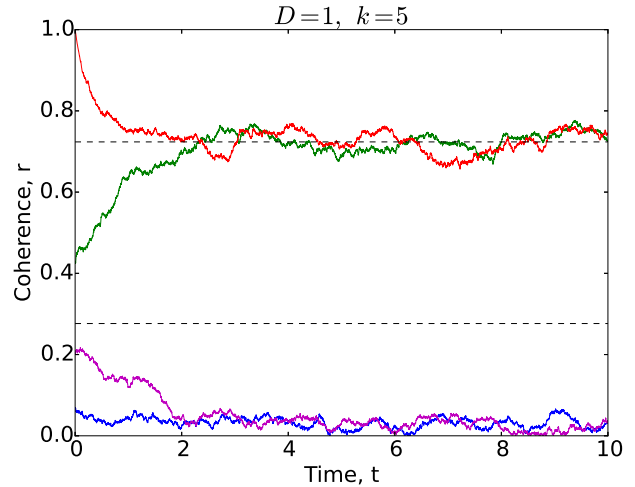}}
  \caption{Temporal evolution of coherence $r(t)$ in population of $N=1000$ identical oscillators with quadratic phase correction rate, $\alpha=k r^2$.
    The diffusion coefficient is $D=1$ and the coupling strength is $k=5>k_c$. 
  Different curves stand for different initial configurations of phases.
  The upper and the lower dashed lines correspond to stable and unstable partially synchronized states of infinite-$N$ population respectively, see Sect. \ref{sec: nonlinear}}
  \label{pic:quadratic}  
  \end{figure}

So far, we have considered  the linear relation between correction rate $\alpha$ and amplitude of the order parameter $r$.
 One may expect more complex and interesting collective behaviour in system with non-linear dependence $\alpha(r)$.
Let us discuss as an illustration the quadratic model $\alpha=kr^2$ focusing on the case of identical oscillators, i.e. $g(\Omega)=\delta(\Omega-\Omega_0)$.
Then, the amplitude equation (\ref{amplitude equation}) reads: 
\begin{equation}
\dot{r}=-Dr-kr^3+kr^2.
\end{equation}
At $k<4D$ the only  time-independent solution is $r=0$ and $r(t)$ decays to zero as $t\to\infty$ for any initial condition $r(0)=r_0$.
However, for $k>4D$ there are three fixed points: $0$ and $r_{\pm}=(1\pm\sqrt{1-4D/k})/2$.
Thus, two partially synchronized branches bifurcate discontinuously from $r=1/2$ at $k_c=4D$, as it is shown in Fig. \ref{pic:nonlinear}.
The incoherence ($r=0$) and the upper synchronous state ($r=r_+$) are stable, and the lower synchronous one ($r=r_-$) is unstable.
This means that the initial configurations with $r_0<r_-$ relax to incoherence as $t\to\infty$,
 while those with  $r_0>r_-$ converge to the synchronized state at long-times. 
The described bistability is clearly observed in numerical simulations of finite-$N$ system, see Fig.~\ref{pic:quadratic}.
Note also that in more general case $\alpha=kr^m$ with $m>1$ the critical coupling is $k_c=Dm^m/(m-1)^{m-1}$ and two partially synchronized branches bifurcate from $r=(m-1)/m$. 
Here again the system is bistable with respect to initial conditions.

For the second particular example $\alpha=kr^{1/2}$ the amplitude equation (\ref{amplitude equation}) yields 
\begin{equation}
\dot{r}=-Dr-kr^{3/2}+kr^{1/2}.
\end{equation}
The incoherent solution $r=0$ is unstable, while the only non-trivial solution $r=[(\sqrt{D^2+4k^2}-D)/(2k)]^2$ is stable and exists for any $k$.
Thus infinitely small coupling leads to spontaneous synchronization of system.
This feature is common for all models $\alpha=kr^m$ with $0\le m<1$.

As it was stated in the previous section, if oscillators have different natural frequencies, there is no closed evolution equation for the order parameter.
Nevertheless, the stationary values of $r$ can be found by substitution $\alpha(r)$ into (\ref{critical coupling}) and solving the resulting self-consistency equation. 
It is easy to see from (\ref{Lorentzian}) that when the frequency diversity is described by a Lorentzian distribution $g(\Omega)$ one should just replace $D$ by $D+\gamma$ in the above formulas for identical oscillators.  
Notice, however, that these arguments do not indicate which of the found branches are stable. 

\section{Conclusion}

In this work, we proposed and investigated a new model of coupled non-identical noisy phase oscillators. 
We assumed that the mutual coupling manifests itself as a tendency of each oscillator to adjust its phase to the
ensemble-averaged value through instantaneous phase shifts which occur at random updating time instants.
Then in the thermodynamic limit the time evolution of the one-oscillator probability density function is governed by the equation (\ref{FPE0}).
The key ingredient of our model is the coherence-dependent rate of updates: the more pronounced the coherence of the current state of population, the more frequently the phase shifts occur.
We mainly concentrated on the simplest linear case when the typical rate of updates is given by the coupling constant multiplied by the coherence.
The distribution of natural frequencies of oscillators was assumed to be unimodal and symmetric.
Our study revealed that the system exhibits the transition to spontaneous synchronization in a fashion similar to the mean-field Kuramoto model. 
Specifically, using the self-consistency arguments for an infinitely large population, we evaluated analytically the critical coupling strength (\ref{critical coupling}) that marks the continuous onset of a partially synchronized solution. 
The same threshold follows from the linear stability analysis of incoherent state.
In the vicinity of the critical value, the linear scaling of the order parameter is found. 
For the sake of illustration, we performed the numerical simulations of large number of oscillators undergoing coherence-induced phase correction.
The analytical predictions for the degree of phase coherence in dependence on the coupling strength turned out to be in good agreement with numerical results, see Fig. (\ref{pic:coherence}).

For homogeneous population of oscillators, we found a closed description of global dynamics, which consists of two uncoupled evolution equations (\ref{phase equation}) and (\ref{amplitude equation}) for the amplitude and phase of the complex order parameter. 
Based on this result, we were able to demonstrate easily the global stability of a partially synchronized state in the system of identical oscillators.

Finally we considered the direct generalization of the model to the case of a nonlinear coupling.
When the relation between the correction rate and the coherence is given by a power law with exponent larger than unity, the synchronization transition is discontinuous (Fig. \ref{pic:nonlinear}) and above the threshold the oscillators exhibit bistability of global activity between incoherence and a partial synchronization (Fig. \ref{pic:quadratic}). 
If the correction rate grows with the coherence slower than linearly, the system shows the non-zero level of synchrony for arbitrary small coupling constant (Fig. \ref{pic:nonlinear}).

The model presented here can be further extended in many ways.
Obviously, the knowledge of the current states of nearby nodes cannot be assumed perfect in a networked applications because of communication and processing time delays and information losses.
From the perspective of the proposed theoretical framework this means that the delta-function in the rhs of our basic equation (\ref{FPE2}) should be replaced by some spread function which depends on the previous states of the system.
Another natural extension of the model includes the short-range interaction effects for regular and complex networks. 
In future studies, it is important also to consider the synchronization properties of system in which the updating time instants of different oscillators are correlated. 

Here we have restricted ourselves to the case of "attractive" phase correction which attempts to mutually synchronize the oscillators. 
There are many practical situations in which the formation of a coherent state in ensemble of oscillatory units is unacceptable  and should be avoided \cite{Louzada_2012,Degesys_2008}. 
One may expect that "repulsive" phase correction ( i.e. the phase shifts to $\psi+\pi$ rather than to $\psi$) is a good strategy to break the undesired synchronization of Kuramoto-like coupled oscillators .


%
%

%

%
%
%
%
%

{}

\end{document}